\documentclass[twocolumn,english,aps,prl]{revtex4}
\usepackage{graphicx}
\begin{document}

\title{Fermi surface topology of Ca$_{1.5}$Sr$_{0.5}$RuO$_{4}$ 
determined by ARPES}

\author{
S.-C. Wang,$^{1}$ H.-B. Yang,$^{1}$ A.K.P. Sekharan,$^{1}$
S. Souma,$^{2}$ H. Matsui,$^{2}$ T. Sato,$^{2}$ T. Takahashi,$^{2}$
Chenxi Lu,$^{3}$ Jiandi Zhang,$^{3}$ R. Jin,$^{4}$ D. Mandrus,$^{4}$ 
E.W. Plummer,$^{4}$
Z. Wang,$^{1}$ and H. Ding$^{1}$
}

\affiliation{
(1) Department of Physics, Boston College, Chestnut Hill, MA 02467\\
(2) Department of Physics, Tohoku University, 980-8578 Sendai, Japan\\
(3) Department of Physics, Florida International University, Miami, FL 
33199\\
(4) Condensed Matter Science Division, Oak Ridge National Laboratory, 
Oak Ridge, TN 37831
}

\begin{abstract}
\noindent We report ARPES results of the Fermi 
surface of Ca$_{1.5}$Sr$_{0.5}$RuO$_{4}$, which is at the boundary of 
magnetic/orbital instability in the phase diagram of the Ca-substituted
Sr ruthenates. Three $t_{2g}$ energy bands and the corresponding Fermi surface 
sheets are observed, which are also present in the Ca-free Sr$_{2}$RuO$_{4}$.
We find that while the Fermi surface 
topology of the $\alpha$, $\beta$ ($d_{yz, zx}$) sheets remains almost the 
same in these two materials, the $\gamma$ ($d_{xy}$) sheet exhibits
a hole-like Fermi surface in Ca$_{1.5}$Sr$_{0.5}$RuO$_{4}$ 
in contrast to being electron-like in Sr$_{2}$RuO$_{4}$. Our observation of 
all three volume conserving Fermi surface sheets clearly demonstrates 
the absence of orbital-selective Mott transition, which was proposed
theoretically to explain the unusual transport and magnetic properties in 
Ca$_{1.5}$Sr$_{0.5}$RuO$_{4}$.
\end{abstract}
\maketitle

The discovery of unconventional superconductivity in Sr$_{2}$RuO$_{4}$
has generated considerable interests in studying the electronic structure of 
the ruthenates \cite{Maeno_Nature}.
Sr$_{2}$RuO$_{4}$, as the only layered perovskite superconductor 
without copper,  has the same crystal structure as the high-$T_{c}$ 
cuprate, La(Sr)$_{2}$CuO$_{4}$.
As in the case of cuprates, Sr$_{2}$RuO$_{4}$ can be regarded 
as being in proximity to a Mott insulator since a complete replacement
of Sr by isovalent Ca leads to an antiferromagnetic (AF) insulator state 
with a moment corresponding to spin \emph{S=1} \cite{Cao}.
It is recently discovered that partial substitution of Sr by isoelectronic Ca 
generates a complex phase diagram for Ca$_{2-x}$Sr$_{x}$RuO$_{4}$ 
\cite{Nakatsuji_PRL00}.  At low Sr concentrations ($0 < x < 0.2$), the 
system is an AF Mott insulator at low temperatures ($T$), a 
metal-insulator transition occurs at a higher $T$.  At higher Sr 
concentrations ($0.2 <x< 0.5$), the system becomes metallic at all $T$ 
with AF correlations at low $T$.  Upon further increasing Sr ($0.5 <x< 
2$), the system becomes a paramagnetic metal, and superconductivity emerges 
at $x$ = 2. The Sr concentration $x_c$ = 0.5 is believed to be at a quantum 
critical point, separating a metallic and orbitally ordered phase ($x < 
x_c$) from the paramagnetic metal ($x > x_c$) \cite{Anisimov}.  The 
spin susceptibility, at the zero temperature limit, is critically 
enhanced at $x_c$ = 0.5, indicating a nearly ferromagnetic (FM) 
instability at this composition \cite{Nakatsuji_PRL00}.
It should be emphasized that since Sr and Ca are isoelectronic, 
Ca-substitution does not change the valence electron numbers in contrast
to carrier doping in the cuprate high-T$_c$ superconductors. The
rich physical phenomena in this series result rather from the changes in the
interplay between electronic correlations and the band structures
induced by the crystal structure changes, and the intriguing possibility 
of ``internal doping'' of selected bands due to the transfer of
valence electrons among the orbitals.

The low-energy excitations of Ca$_{2-x}$Sr$_{x}$RuO$_{4}$ are believed to 
originate from the hybridization between Ru-4$d$ $t_{2g}$ and 
O-2$p$ orbitals. It is generally accepted that, at the one end of the 
phase diagram, Sr$_{2}$RuO$_{4}$ has four valence electrons evenly 
distributed in the three $t_{2g}$ orbitals,
forming three energy bands $\alpha$, $\beta$ (due to the mixing of $d_{yz}$ 
and $d_{zx}$ states) and $\gamma$ ($d_{xy}$). The electron
occupancy has the fractions ($n_{(\alpha,\beta)}$, 
$n_{\gamma}$)=($\frac{8}{3}$, $\frac{4}{3}$). At the other end, 
Ca$_{2}$RuO$_{4}$'s electron
distribution is believed to be (2,2), creating a Mott localized AF
ground state at low $T$ \cite{Anisimov}. However, the evolution
of the electron distribution and band structure in partially substituted
Ca$_{2-x}$Sr$_{x}$RuO$_{4}$ is not fully understood yet. In particular, 
the knowledge of the electronic structure of Ca$_{1.5}$Sr$_{0.5}$RuO$_{4}$ 
is important in understanding the alleged critical behavior and the 
transition to the Mott insulator.

It is proposed by Anisimov \emph{et al}. \cite{Anisimov}, based on a 
Non-Crossing Approximation (NCA) calculation within Dynamical Mean 
Field Theory (DMFT), that the valence electron distribution becomes
(3,1) at $x_{c}$. They further proposed that one of the $\alpha$ and
$\beta$ bands hosts 2 of the 3 electrons and is thus completely filled
and band-insulating while the other, half-filled band becomes 
Mott-localized with spin-$\frac{1}{2}$ local moment 
due to the narrow bandwidth relative to the
Coulomb energy (Hubbard-U). The half-filled $\gamma$ band, however, 
remains itinerant due to its wider bandwidth, resulting in a metallic phase 
consistent with the experiments 
\cite{Nakatsuji_PRL00,Nakatsuji_PRB}.
This proposal of an apparent orbital-selective Mott transition (OSMT)
is of general interests to multi-band correlated systems such
as transition metal oxides and the heavy fermion compounds. However,
the (non)existence of OSMT, particularly in real materials, has been
a much debated issue \cite{Liebsch,Koga,Liebsch_new}. 
From a quantum mechanical point of view, localized and extended states
cannot coexist at the same energy unless quantum tunneling (mixing) between
these states are strictly forbidden by symmetry considerations.
Indeed, a different Quantum Monte Carlo (QMC) calculation within the
DMFT by Liebsch \cite{Liebsch} suggests a common metal-insulator 
transition for all three $t_{2g}$ bands at a same critical correlation 
$U_c$. The debate on this issue continues with two recent theoretical papers 
reaching opposite conclusions \cite{Koga,Liebsch_new}. 
However, there has been no experimental test
that such OSMT exists in degenerate $d$-electron systems with small 
bandwidth difference among orbitals.

Angle-resolved photoelectron spectroscopy (ARPES) is a suitable 
experimental technique to study this problem because of its ability to
determine band dispersion and the Fermi surface in the momentum space. 
The technique of de Hass-van Alphen (dHvA) is another commonly used 
method for measuring the Fermi surface. It has played a crucial role in 
mapping the three Fermi sheets in the undoped Sr$_{2}$RuO$_{4}$. 
However, it is difficult to use dHvA to probe Ca-doped ruthenate due to 
disorders in the doped materials. Therefore ARPES is perhaps a unique
experimental tool to resolve this controversy, as suggested originally
by Anisimov \emph{et al} \cite{Anisimov}.  We have measured 
extensively Ca$_{1.5}$Sr$_{0.5}$RuO$_{4}$ single crystals and performed
a comparative study to Sr$_{2}$RuO$_{4}$, as reported below.

High-quality Ca$_{2-x}$Sr$_{x}$RuO$_{4}$ single crystals are prepared
by the floating zone method \cite{Jin}. ARPES experiments are performed 
at the Synchrotron
Radiation Center, Wisconsin using undulator beamlines (U1 NIM and PGM) 
at different photon energies (10 to 32 $eV$). Samples are cleaved
\emph{in situ} and measured at $T$ = 40 $K$ in a vacuum better than 
$8\times10^{-11}\; Torr$. A Scienta analyzer capable of multi-angle
detection is used with energy resolution of 10 - 20 $meV$, and momentum
resolution of $\sim0.02$\AA$^{-1}$. Samples are stable and show no sign 
of degradation during a typical measurement period of 12 hours.

Similar to Sr$_{2}$RuO$_{4}$, Ca$_{1.5}$Sr$_{0.5}$RuO$_{4}$ is easy to 
cleave and usually has a good (001) surface. Both materials exhibit clear 
LEED patterns for their cleaved surfaces, as shown in Figs.~1(a)-(b). 
The brighter LEED spots form a square lattice, corresponding to the 2D 
RuO$_2$ lattice. The additional faint spots, which appear in the middle 
of four bright spots, is caused by the rotation of RuO$_6$-octahedra 
along the $c$-axis. In Sr$_{2}$RuO$_{4}$, this rotation is caused by a 
surface reconstruction driven by a soft phonon mode \cite{Matzdorf}. 
However, this rotation is observed to exist in the bulk of 
Ca$_{2-x}$Sr$_{x}$RuO$_{4}$ when $x < 1.5$ \cite{Friedt}. For 
Ca$_{1.5}$Sr$_{0.5}$RuO$_{4}$, the rotation angle in the bulk is 
about 12$^\circ$ \cite{Friedt}. In comparison, the rotation angle on 
the surface obtained by LEED analysis is about 11$^\circ$, indicating 
similar crystal structures between the bulk and surface. Therefore we 
expect ARPES results on Ca$_{1.5}$Sr$_{0.5}$RuO$_{4}$ are bulk 
representative.

Due to the rotation of RuO$_6$-octahedra, the 2D Brillouin zone (BZ) 
becomes a $\sqrt{2}$x$\sqrt{2}$ lattice rotated 45$^\circ$ 
with respect to the original 1x1 square lattice. Nevertheless, in this 
paper, we still use the 1x1 square lattice to discuss the band structure
for convenience in comparison.  The main effect of this rotation on
the band structure is a band folding with respect to the new zone 
boundaries.
In Sr$_{2}$RuO$_{4}$, the surface states due to the lattice rotation 
can be greatly suppressed by varying photon energy or aging the surface 
in ARPES experiments.  Thus bulk-representative band structure and Fermi 
surface topology, similar to the ones predicted by band calculations 
\cite{Oguchi,Singh} and observed by dHvA
measurement \cite{Mackenzie}, have been observed by ARPES on 
Sr$_{2}$RuO$_{4}$ \cite{Damascelli,Ding_PhysicaC,KMShen}.

\begin{figure}
\includegraphics[width=3.6in]{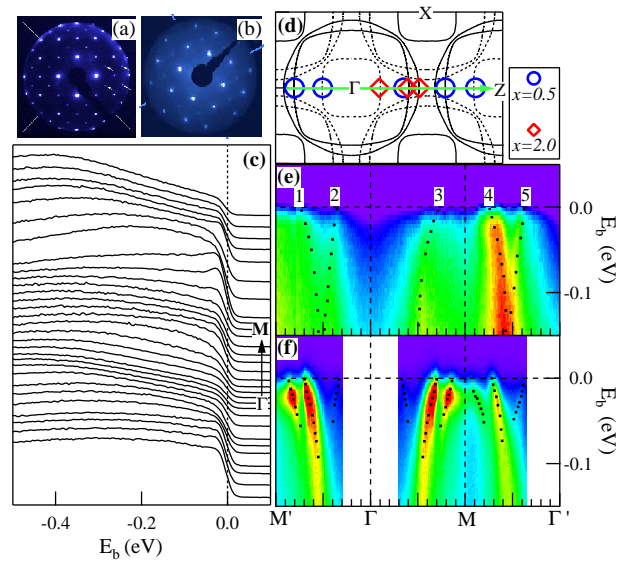}
\vspace{-20pt}
\caption{LEED patterns for (a) Ca$_{1.5}$Sr$_{0.5}$RuO$_{4}$, and (b) 
Sr$_{2}$RuO$_{4}$. (c) EDCs 
of Ca$_{1.5}$Sr$_{0.5}$RuO$_{4}$ along $\Gamma$-$M$ using 32-$eV$ photons.
(d) Measurement locations (green line) in extended BZs, with the 
calculated FS (solid lines) of Sr$_2$RuO$_4$ and the ``image'' FS 
(dashed lines). FS crossing points are plotted for 
Ca$_{1.5}$Sr$_{0.5}$RuO$_{4}$ (blue circles) and Sr$_2$RuO$_4$ (red 
diamonds) extracted from panels below.
$E$-$k$ intensity plots for (e) Ca$_{1.5}$Sr$_{0.5}$RuO$_{4}$,
and (f) Sr$_2$RuO$_4$. The dots are guides for band dispersion.}
\vspace{-10pt}
\end{figure}

In the following we focus on the electronic structure near $E_F$, 
which determines the low energy properties of this material. In Fig.~1(c) we 
plot the energy distribution curves (EDCs) of Ca$_{1.5}$Sr$_{0.5}$RuO$_{4}$ 
along the (0,0)-($\pi$,0) ($\Gamma$-$M$, the Ru-O bond) direction 
over several BZs. One can clearly observe band dispersion and Fermi 
surface crossings (FSCs). 
To see the dispersion and FSCs more clearly, we display the corresponding 
$E$-$k$ intensity plot in Fig.~1(e).
We can identify five dispersive bands, as marked in Fig.~1(e). The 
strongest feature is \#4 in the 2nd BZ, which has equivalent bands in other 
BZs, such as \#1 and \#3 whose $k$-locations are 
shifted by a reciprocal lattice vector $G$. In comparison to band 
calculations \cite{Oguchi,Singh}, dHvA measurements
\cite{Mackenzie}, and ARPES results 
\cite{Damascelli,Ding_PhysicaC,KMShen} on Sr$_{2}$RuO$_{4}$, we find 
that these bands (\#1, 3, and 4) match almost perfectly to the $\beta$ 
band. As an illustration, we plot the extracted FSCs of 
Ca$_{1.5}$Sr$_{0.5}$RuO$_{4}$ in Fig.~1(d) where the calculated 
Sr$_{2}$RuO$_{4}$ FS sheets (solid lines) are shown. In Fig.~1(d) we 
also plot the FSCs of Sr$_{2}$RuO$_{4}$ extracted from Fig.~1(f).

Fig.~1(f) displays an $E$-$k$ intensity plot along $\Gamma$-$M$ for 
Sr$_{2}$RuO$_{4}$. In addition to the $\beta$ band, the $\gamma$ band 
is observed near $M$. However, no such a band is visible along 
$\Gamma$-$M$ in Ca$_{1.5}$Sr$_{0.5}$RuO$_{4}$, as shown in Fig.~1(e). 
One may argue that the observed bands (\#1, 3, and 4) belong to the 
$\gamma$ band that shrinks its FS area upon Ca substitution. This 
scenario is unlikely due to the observation of a reversed band 
dispersion, such as \#5 and \#2 in Fig.~1(e). We attribute this 
reversed band dispersion to a folded $\alpha$ band caused by the rotation of 
RuO$_6$-octahedra in the bulk of Ca$_{1.5}$Sr$_{0.5}$RuO$_{4}$ 
\cite{Friedt}. This band folding introduces the so-called ``image'' FS, as 
shown in Fig.~1(d) (dashed lines). The FSCs of the reversed bands \#5 
and \#2 match well with the predicted  ``image'' of $\alpha$ FS. Note a 
similar band folding is also observed in Sr$_{2}$RuO$_{4}$, which is 
due to the rotation of RuO$_6$-octahedra on the surface, as discussed 
before.

\begin{figure}
\includegraphics[width=3.4in]{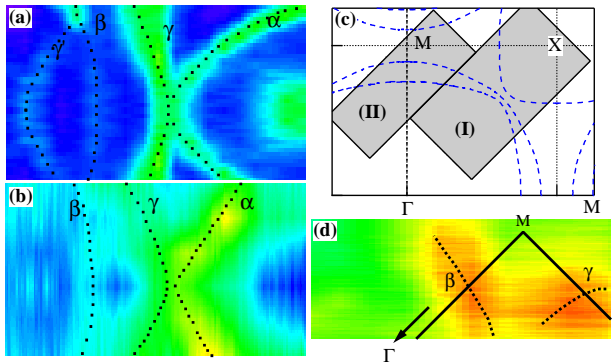}
\vspace{-30pt}
\caption{2D plots of intensity integrated over a small energy region of $E_F$ 
($\pm$ 20 meV) for: 
(a) Sr$_{2}$RuO$_{4}$ in the rectangular box (I) shown in (c); 
(b) Ca$_{1.5}$Sr$_{0.5}$RuO$_{4}$ in  the same box (I); 
(d) Ca$_{1.5}$Sr$_{0.5}$RuO$_{4}$ in the box (II). 
(c) Locations of measurement in the BZ.}
\vspace{-10pt}
\end{figure}

As seen in the band calculation, the three Fermi surfaces are adjacent 
in the vicinity of (2/3$\pi$, 2/3$\pi$). Thus this location is a good 
place to observe all the three FS and their relative positions. This is 
clearly demonstrated in the case of Sr$_{2}$RuO$_{4}$, as shown in Fig.~2(a), 
where the near-$E_F$ ($\pm$ 20 meV) intensity is plotted in a 2D 
$k$-region indicated by the shaded rectangular area (I) in the BZ shown 
in Fig. ~2(c). One can clearly observe the three ($\alpha$, $\beta$, 
$\gamma$) FS sheets in Fig.~2(a), although the intensity of the $\beta$ 
FS is much weaker along $\Gamma$-$X$. This is most likely due to the 
selection rule of ARPES.  While the $\alpha$ and $\beta$ FS sheets are 
well separated due to orbital hybridization between them, the $\gamma$ 
FS ``touches'' the $\alpha$ FS along $\Gamma$-$X$, reflecting the 
non-mixing nature between the $xy$ and $yz(zx)$-orbitals, as predicted 
based on a symmetry argument \cite{Rice}. In addition, we observe the 
folded $\gamma^{\prime}$ FS and a small FS pocket at the $X$ point, 
which is believed to be a result of the rotation of RuO$_6$-octahedra 
on the surface of Sr$_{2}$RuO$_{4}$ \cite{KMShen}.

For Ca$_{1.5}$Sr$_{0.5}$RuO$_{4}$, as shown in Fig.~2(b), one can still 
observe the three FS sheets, although their intensity is weaker 
than in Sr$_{2}$RuO$_{4}$. The reduction in spectral intensity at $E_F$ 
is commonly observed in doped correlated systems. This decoherence 
phenomenon is believed to be caused by correlation and disorder effects. 
Among the three FSs, the $\alpha$ FS is the most visible. A ``faint'' 
FS, which ``touches'' the $\alpha$ FS along $\Gamma$-$X$, should belong 
to the $\gamma$ band due to its non- or weak-mixing with the $\alpha$ 
band. Another less visible FS, which is further separated from the 
$\alpha$ FS, is naturally assigned to the $\beta$ band.

As discussed above, we do not observe the $\gamma$ FS crossing along 
$\Gamma$-$M$. It is possible that the $\gamma$ FS changes its topology 
from electron-like centered at $\Gamma$ to hole-like centered at $X$. 
To check this, we measured the ARPES spectra in the vicinity of $M$, 
indicated by the rectangular box (II) in the BZ shown in 
Fig.~2(c). From the plot of the near-$E_F$ intensity in Fig.~2(d), one 
can observe two FS sections: the one intersecting with $\Gamma$-$M$ 
belongs to the $\beta$ FS, and the other intersecting with $M$-$Y$ 
should be the $\gamma$ FS which becomes hole-like by enclosing the $X$ 
point.

\begin{figure}
\includegraphics[width=2.2in]{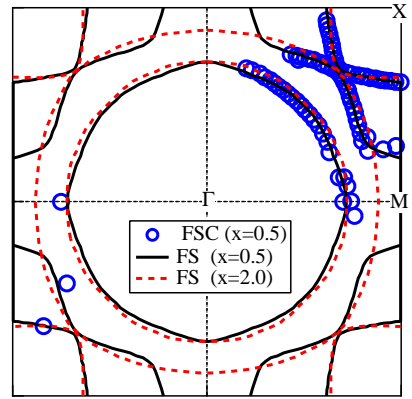}
\vspace{-10pt}
\caption{Measured FS crossing points (blue circles) and derived Fermi surfaces 
(solid black lines) in Ca$_{1.5}$Sr$_{0.5}$RuO$_{4}$. For comparison, extracted 
FSs (red dashed lines) in Sr$_{2}$RuO$_{4}$ are also plotted . Note the folded 
``image'' FSs are not plotted for clarity.
}
\vspace{-10pt}
\end{figure}

We summarize our ARPES results of Ca$_{1.5}$Sr$_{0.5}$RuO$_{4}$ by 
plotting its measured Fermi surface in Fig.~3. The blue circles are the FS 
crossing points directly obtained from ARPES spectra. The black solid 
lines are derived Fermi surfaces based on the measurement and trivial 
symmetry operations. For the purpose of comparison, we also plot in 
Fig.~3 the Fermi surfaces (red dashed lines) of Sr$_{2}$RuO$_{4}$ 
determined from our ARPES experiment. It is clear from Fig.~3 that all 
the three FS sheets, with similar topology, are observed in both 
Ca$_{1.5}$Sr$_{0.5}$RuO$_{4}$ and Sr$_{2}$RuO$_{4}$ with the significant
difference that the $\gamma$ FS changes from electron-like in 
Sr$_{2}$RuO$_{4}$ to hole-like in Ca$_{1.5}$Sr$_{0.5}$RuO$_{4}$. This 
change is consistent with the LDA band calculations 
\cite{Singh,Fang,Fang_new} for the rotated crystal structure.  A 
similar change of the $\gamma$ FS topology is also observed for the 
surface state in Sr$_{2}$RuO$_{4}$, which has the similar structure 
rotation on the surface \cite{KMShen}.

From the determined FS topology in Fig.~3, we can extract the values of 
the Fermi vector ($k_F$) for the three Fermi surface sheets along the high symmetry 
lines, as listed in Table 1. We note that the $k_F$ values are similar 
to the ones obtained from dHvA measurement \cite{Mackenzie} for 
Sr$_{2}$RuO$_{4}$. We also determined the occupied area for the three FS 
sheets, and find no appreciable differences between the two materials.
These observations suggest that there is 
no significant electron transfer among the three orbitals at this 
Ca doping level, in contrast to the scenario proposed by Anisimov \emph{et al}
\cite{Anisimov}.
The total occupied area of the two materials are close 
to 2, indicative of 4 electrons per unit cell, satisfying Luttinger theorem.

\begin{table}[htdp]
\begin{center}
\begin{tabular}{|c|c|c|}
\hline
 	& Ca$_{1.5}$Sr$_{0.5}$RuO$_4$ &  Sr$_2$RuO$_4$ \\
\hline
$k_F(x)$ ($\Gamma$-M) & $\beta$(0.72) & $\beta$(0.72), $\gamma$(0.88) \\
\hline
$k_F(x=y)$ ($\Gamma$-X) & $\alpha/\gamma$(0.67), $\beta$(0.50) & $\alpha/\gamma$(0.67), $\beta$(0.51) \\
\hline
$k_F(y)$ (M-X) & $\alpha$(0.62), $\gamma$(0.22) & $\alpha$(0.64) \\
\hline
FS Area & $\alpha$(0.86), $\beta$(0.38), & $\alpha$(0.86),$\beta$(0.41), \\
         & $\gamma$(0.68) & $\gamma$(0.64) \\
\hline
Total Area & 1.92 & 1.91 \\
\hline
\end{tabular}
\end{center}
\vspace{-10pt}
\caption{$k_F$ of FS crossing points along high symmetry lines, and the occupied 
FS area in Ca$_{1.5}$Sr$_{0.5}$RuO$_{4}$ and Sr$_{2}$RuO$_{4}$.
The experimental uncertainty of $k_F$ is $\pm 0.04$. The unit of $k_F$ is $\pi /a$, where $a$=3.76 (3.86)\AA\, for Ca$_{1.5}$Sr$_{0.5}$RuO$_4$ (Sr$_2$RuO$_4$).} 
\label{default}
\end{table}

The observation of all the three Fermi surface sheets in 
Ca$_{1.5}$Sr$_{0.5}$RuO$_{4}$ and the lack of significant inter-orbital
electron transfer clearly demonstrate the absence of 
OSMT in this material. The proposal of 
OSMT at $x$ = 0.5 was originally motivated by the experimental 
observation of spin-$\frac{1}{2}$ local moment \cite{Anisimov}. 
The evidence for the $\frac{1}{2}$ local moment comes from fitting the 
magnetic susceptibility in Ca$_{1.5}$Sr$_{0.5}$RuO$_{4}$ with the 
Curie-Weiss form. Our observation 
of three itinerant orbitals appears to be inconsistent with the existence
of such localized moments, although more exotic mechanisms may account for
(to a certain degree) the coexistence of the local moment and itinerant 
electrons. We note a recent polarized neutron diffraction experiment for $x$ = 0.5 \cite{Gukasov} suggesting that the dominant magnetization distribution originates from 
the $xy$ orbital rather than the $yz$/$zx$ orbitals predicted by the 
OSMT theory.
Nevertheless, the question of what type of Mott transition takes place upon further 
Ca doping remains. Whether or not it goes through OSMT is still an open issue, 
and requires more experimental efforts. A recent high-filed study \cite{Nakatsuji_PRL03} reports the saturation magnetic moment in Ca$_{1.8}$Sr$_{0.5}$RuO$_{4}$ ($x$ = 0.2) 
is close to the effective magneton of 1 $\mu_B$ that is  expected from a system with 
a local moment of $S$ = 1/2, indicating a possible OSMT at $x$ = 0.2.

In summary, we observe all the three $t_{2g}$ bands and the 
corresponding Fermi surface sheets in both 
Ca$_{1.5}$Sr$_{0.5}$RuO$_{4}$ and Sr$_{2}$RuO$_{4}$ by ARPES 
experiment. The most significant change of the Fermi surface topology is the 
$\gamma$ Fermi surface which becomes hole-like in 
Ca$_{1.5}$Sr$_{0.5}$RuO$_{4}$ near the $M$ point,
while being electron-like in Sr$_{2}$RuO$_{4}$. This is likely caused 
by the rotation of RuO$_6$-octahedra along the $c$-axis in the bulk upon 
the Ca substitution. Together with our observation that the electron
filling fractions are conserved approximately within each of the $t_{2g}$ orbitals
when compared to Sr$_{2}$RuO$_{4}$, we conclude that the 
orbital-selective Mott transition is absent at $x=0.5$ and
call for further understanding of the electronic structure, correlation
effects, transport and magnetic properties in Ca-doped ruthenates.

We thank S. Gorovikov, C. Gundelach, and H. Hochst for technical 
support in synchrotron experiments, Z. Fang, A. Liebsch, A. Millis, 
T.M. Rice, and D. Xi for useful discussions and suggestions.
This work is supported by NSF DMR-0072205, NSF DMR-0072998, 
DOE DE-FG02-99ER45747, Petroleum 
Research Fund, Sloan Foundation, and the MEXT of Japan.
The Synchrotron Radiation Center is supported by NSF DMR-0084402. Oak 
Ridge National laboratory is managed by UT-Battelle, LLC, for the U.S. 
Department of Energy under contract
DE-AC05-00OR22725.

\end{document}